\documentclass[a4paper]{article}

\usepackage[group-separator={,}]{siunitx}
\usepackage{INTERSPEECH2022}

\usepackage{textcomp}
\newcommand{\textapprox}{\raisebox{0.5ex}{\texttildelow}}

\usepackage[dvipsnames]{xcolor}
\usepackage[normalem]{ulem}

\usepackage{subcaption}

\usepackage{booktabs}
\usepackage[textfont=it,tableposition=top]{caption}

\title{Discrete Acoustic Space for an Efficient Sampling in Neural Text-To-Speech}
\name{Marek Strong, Jonas Rohnke, Antonio Bonafonte, Mateusz Łajszczak, Trevor Wood}
\address{Amazon, Alexa AI}
  
\email{\{strelecm, rohnj, bonafont, mateuszl, trevowoo\}@amazon.com}

\begin{document}

\maketitle
\begin{abstract}
We present a Split Vector Quantized Variational Autoencoder (SVQ-VAE) architecture using a split vector quantizer for NTTS, as an enhancement to the well-known Variational Autoencoder (VAE) and Vector Quantized Variational Autoencoder (VQ-VAE) architectures. Compared to these previous architectures, our proposed model retains the benefits of using an utterance-level bottleneck, while keeping significant representation power and a discretized latent space small enough for efficient prediction from text. We train the model on recordings in the expressive task-oriented dialogues  domain and show that SVQ-VAE achieves a statistically significant improvement in naturalness over the VAE and VQ-VAE models. Furthermore, we demonstrate that the SVQ-VAE latent acoustic space is predictable from text, reducing the gap between the standard constant vector synthesis and vocoded recordings by 32\%.
\end{abstract}
\noindent\textbf{Index Terms}: prosody control, expressive speech synthesis, VQ-VAE, NTTS


\section{Introduction}
\label{sec:intro}

Generating speech for complex scenarios such as multi-turn dialogues requires modeling the rich prosodic aspects of speech to capture the natural variability of human conversation and assign the appropriate conversational style. Recent advances in TTS \cite{shen2018natural,prateek2019other} have allowed for the generation of high-quality speech but suffer from a lack of appropriate variation. As a result, the generated audio often sounds monotonous and unnatural.


One of the popular techniques for improving prosodic variation is based on extracting prosodic features from ground-truth data via Variational Auto-Encoders (VAE) \cite{akuzawa2018expressive,hsu2018hierarchical, karanasou2021learned,zhang2019learning}. These methods use reference mel-spectrograms during training but have to provide the acoustic features from a different source during inference, as mel-spectrograms are unavailable. For example, Ezzerg et al. \cite{ezzerg2021enhancing} computed a centroid (mean) over training data to represent the acoustic space. While using a constant vector provides a simple solution, this approach often results in monotonous speech. Akuzawa et al. \cite{akuzawa2018expressive} experimented with random sampling from the VAE’s prior, and Hodari et al. \cite{hodari2019using} proposed sampling focusing on low-probability regions of the prior distribution to produce speech with varied intonation. These methods introduce variation but can lead to unnatural speech due to the randomness in sampling. Karlapati et al. \cite{karlapati2021prosodic} introduced more sophisticated samplers that combine BERT and Graph samplers to predict Gaussian parameters from which they sample the final acoustic vector. Tyagi et al.\cite{tyagi2020dynamic} applied selection-based strategies to search a specific set of utterances for the most appropriate match based on syntactic structure.

All of these approaches, however, suffer from the fact that the underlying acoustic latent space often represents both high-level (e.g., general speaking style) and low-level (e.g., prosodic patterns) aspects of speech. This is not desirable for synthesis with constant acoustic vectors (e.g., centroids) because they are likely to encode noise and unique characteristics of particular utterances, resulting in instabilities and inconsistent style generation. Moreover, predicting such a multi-modal distribution from text is difficult.

To simplify the acoustic latent space for efficient sampling, we apply Vector Quantized-Variational AutoEncoder (VQ-VAE), which has been shown to learn a high-level abstract space of features \cite{oord2017neural,razavi2019generating}. In this work, we take inspiration from \cite{kaiser2018fast} and present a modified version of VQ-VAE, called Split Vector Quantized Variational
Autoencoder (SVQ-VAE), that partitions the latent vector into several splits that are each discretized independently. This allows the model to keep a smaller number of codes while increasing the number of information bits that can be encoded. This architecture gives the model significant representation power while keeping the discretized latent space small enough for efficient prediction from text. 

While our work focuses on utterance-level prosody representations, more fine-grained representations have been proposed as well. A recent study by Hodari et al. \cite{hodari2021camp} applies a two-stage training framework for prosody modeling, where word-level prosody representation is first learned separately and then predicted from text for TTS. Hono et al. \cite{hono2020hierarchical} proposed a hierarchical approach that captures three different time resolutions (utterance-level, phrase-level, and word-level), where the prediction of each level is conditioned on latent variables from the coarser-level. Our work demonstrates that significant improvements can be achieved by using simpler utterance-level representations. Sun et al. \cite{sun2020fully} also used a hierarchical VAE model to disentangle prosodic features on different levels. However, their work concerns the control of the prosody style rather than predictability from context. Similarly, Sun et al. \cite{sun2020generating} investigates fine-grained architectures with auto-regressive prosody priors, but they focus on random sampling and diverse prosodic variations for ASR.

This work makes two contributions: 1) We present an SVQ-VAE model for TTS and show that it outperforms its VQ-VAE and VAE counterparts on constant-vector synthesis. 2) We demonstrate that the SVQ-VAE acoustic space can be predicted from BERT's word-piece embeddings, reducing the gap between centroid synthesis and vocoded recordings by 32\%.

\begin{figure}
\centering
\includegraphics[width=1.\linewidth]{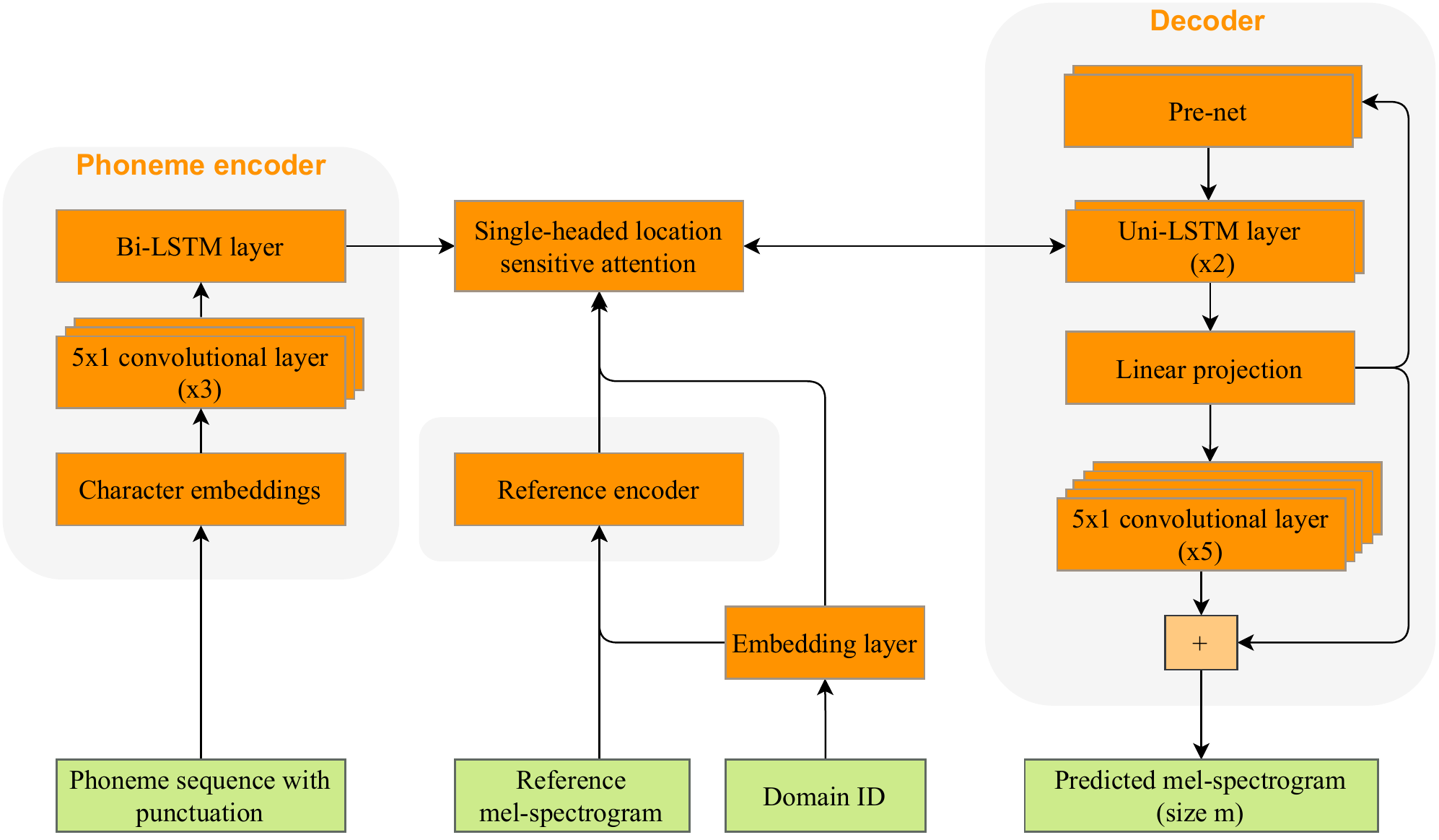}
\caption{NTTS model architecture}
\label{fig:arch_ntts}
\end{figure}


\section{Model}
\label{sec:format}

Our TTS system is based on a Tacotron2-like \cite{shen2018natural} architecture with an additional module for learning high-level speaking styles in an unsupervised manner. As shown in Figure \ref{fig:arch_ntts}, this model takes phonemes and domain ID as inputs, where domain ID is used to identify each of the pre-defined subsets of data (e.g., neutral, emotions, dialogues, etc.). The output of the model is acoustic features (mel-spectrograms), which are converted into audio waveforms using a vocoder \cite{jiao2021universal,shen2018natural}. This section shows the details of our proposed model, comparing VAE, VQ-VAE, and SVQ-VAE reference encoders. Finally, in section~\ref{sec:pred_acoustic_space}, we describe the architecture to predict the latent codes from the input text.

\subsection{Acoustic model}

The acoustic model consists of the phoneme encoder, reference encoder, domain encoder, single-headed location-sensitive attention, and decoder.

In order to synthesize speech, the three encoders of the acoustic model first encode all available input features. The phoneme encoder uses stacked convolutional layers and a Bi-LSTM layer to encode the input phoneme sequence. The reference encoder \cite{skerry2018towards} takes mel-spectrograms and applies a bottleneck to model utterance-level speech styles (see sections \ref{sec:vae}-\ref{sec:svq}). The domain encoder uses an embedding lookup layer on a one-hot vector representing each of the pre-defined domains in our training dataset. The encoder outputs are concatenated and fed to an attention layer, which summarizes the entire encoded sequence as a fixed-length context vector for each decoder output step. The layer uses the location-sensitive attention mechanism from \cite{chorowski2015attention}, which uses cumulative attention weights to prevent the model from skipping or repeating segments of speech.

Finally, the decoder predicts 5 frames at each decoding step. Each generated frame has a dimensionality of 80 and represents 50\,ms of speech with an overlap of 12.5\,ms.

\subsection{VAE reference encoder}
\label{sec:vae}

\begin{figure}[hbt!]
\centering
\begin{subfigure}{.14\textwidth}
  \centering
  \includegraphics[width=1.\linewidth]{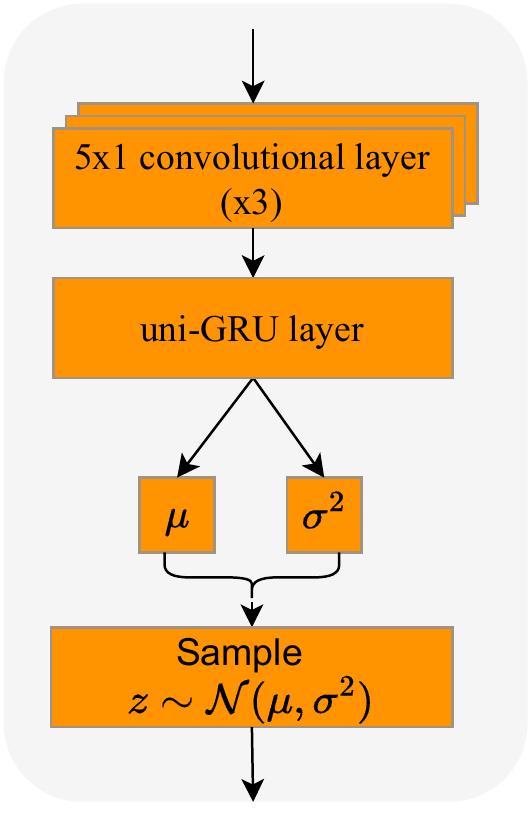}
  \caption{VAE}
  \label{fig:arch_ref_encoders_vae}
\end{subfigure}
\quad
\begin{subfigure}{.14\textwidth}
  \centering
  \includegraphics[width=1.\linewidth]{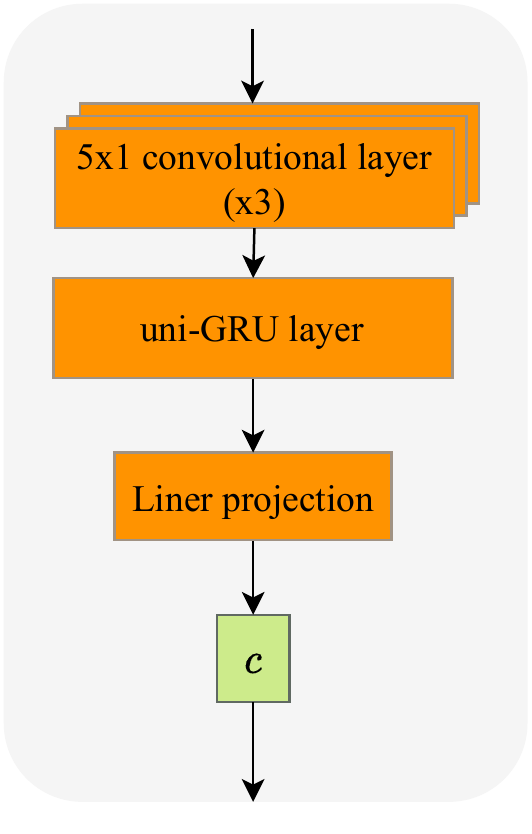}
  \caption{VQ-VAE}
  \label{fig:arch_ref_encoders_vq}
\end{subfigure}
\quad
\begin{subfigure}{.14\textwidth}
  \centering
  \includegraphics[width=1.\linewidth]{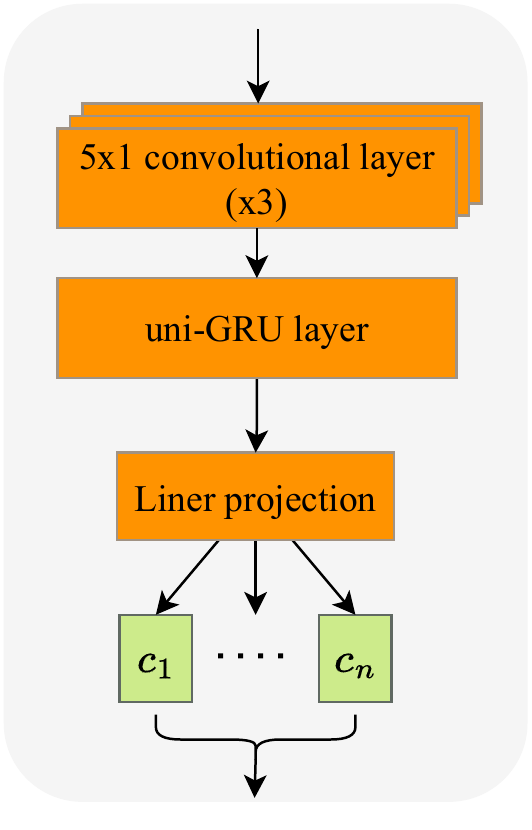}
  \caption{SVQ-VAE}
  \label{fig:arch_ref_encoders_svq}
\end{subfigure}
\caption{A detailed view into the three reference encoders used for centroid-based synthesis. The SVQ-VAE encoder was also used for experiments with predicted synthesis.}

\label{fig:arch_ref_encoders}

\end{figure}

The baseline model uses a variational reference encoder, as shown in Figure \ref{fig:arch_ref_encoders_vae}. During training, the encoder summarizes the mel-spectrogram input into a single 128-dimensional vector and predicts the mean and standard deviation of a Gaussian distribution - $\mu$ and $\sigma$. We assume a prior distribution $p_{\theta}(z) = N(0, I)$, where $\theta$ represents the parameters of the reference encoder and $z$ is the produced latent variable. The variable $z$ is sampled using the reparameterization trick \cite{kingma2013auto}, and the model is trained to maximize the evidence lower bound (ELBO) \cite{kingma2019introduction}. Similar to \cite{zhang2019learning}, we apply KLD loss annealing and scheduling to avoid posterior collapse.

\subsection{VQ-VAE reference encoder}
\label{sec:vq}

While the continuous nature of the VAE latent space provides significant representation power, it also results in capturing properties of speech that generally do not need to be modeled with such high fidelity. Consequently, we use VQ-VAE (Figure \ref{fig:arch_ref_encoders_vq}) to apply a strong bottleneck on these acoustic features. 

As introduced by Oord et al. \cite{oord2017neural}, the VQ-VAE encoder uses vector quantization in order to obtain a discrete latent representation of the continuous space. More specifically, our VQ-VAE encoder first summarizes the mel-spectrogram input into a single 128-dimensional vector. It then projects it into a codebook space C, which consists of K codebook vectors (or codes) $c_{1}, c_{2}, \ldots, c_{K}$, each of dimensionality $D$. During the forward pass, VQ-VAE computes the L2-normalized distance between the encoder output and all codebook vectors, choosing the one with the minimum distance. Finally, this vector is passed to the decoder.

The choice of the codebook size K and the dimensionality of codes D have a major impact on the acoustic representation of the input. Small codebooks allow for an easy interpretation of codes and clear separation of styles, but we have experimentally observed that the overall speech quality is reduced. In our experiments, a codebook size of at least $512$ is necessary to produce good quality speech. While the speech quality improves further as we increase the size of the codebook, using too many codes can affect generalization as there are only $57,975$ utterances in our training dataset. In our internal evaluations, we achieved the best results with $8,192$ codes, which is the value used for experiments in this paper.

In the context of VQ-VAE, a codebook collapse is the equivalent of a posterior collapse in VAE. It occurs when a portion of the codes become underutilized, causing the approximate posterior $q_{\theta}(z|x)$ to be supported by only a handful of codes. To address this problem, we use random restarts as Dhariwal et al. \cite{dhariwal2020jukebox}.

\subsection{SVQ-VAE reference encoder}
\label{sec:svq}

The empirical observations of the previous section regarding optimal codebook size can be better understood if we consider the information capacity required to learn useful audio representations. Let us assume we want to optimally represent global attributes of audio (e.g. pace, intonation, emotion) in a discrete VQ-VAE space, and that these attributes can be represented by a random variable with $x_{1},x_{2},\ldots,x_{F}$ features. If we assume that each feature has $y_{1},y_{2},\ldots,y_{N}$ possible values, in order to capture all combinations of these values in the VQ-VAE space, we need a codebook with $F^{N}$ codes. Thus, in domains with high acoustic diversity, we need large codebooks for good coverage of possible speech variations. This presents several challenges. First, the VQ-VAE approach becomes impractical as the memory footprint gets too high for large codebooks. Second, a codebook with a large number of codes is more likely to result in a codebook collapse. Frequently used codes receive a stronger signal than other codes, leading to a further increase in their usage and, ultimately, only a small subset of codes being utilized. Lastly, increasing the number of codes can have a negative impact on the prediction of codes from text, increasing the total number of targets and thus making them more sparse.

In order to be able to encode more information while keeping the number of codes in a codebook reasonably low, we apply a split vector quantizer \cite{paliwal1993efficient,kaiser2018fast} shown in Figure \ref{fig:arch_ref_encoders_svq}. It encodes different parts of the latent vector individually, using separate codebooks. More specifically, our SVQ-VAE reference encoder uses $S$ codebooks $C_{1},C_{2},\ldots,C_{S}$. The output vector of the GRU is partitioned into $S$ splits, and each split is passed into a dedicated codebook. Similar to the standard VQ-VAE, a code is selected by performing a nearest-neighbor lookup, resulting in $S$ selected codebook vectors. These vectors are concatenated to obtain the final acoustic representation of the input. Note that for $S=1$, we get the original VQ-VAE reference encoder. Comparing VQ-VAE with SVQ-VAE, the former can encode only $\log_{2}K$ information bits for a codebook size of $K$ while the latter can encode $S\log_{2}K$.


We conducted an extensive hyper-parameter search on our development dataset and achieved the best results with $8$ splits, $1,024$ codes per codebook, and codebook dimensionality of $8$.  Therefore, these parameters are used throughout all experiments presented.

\begin{figure}[hbt!]
\centering
\includegraphics[width=0.9\linewidth]{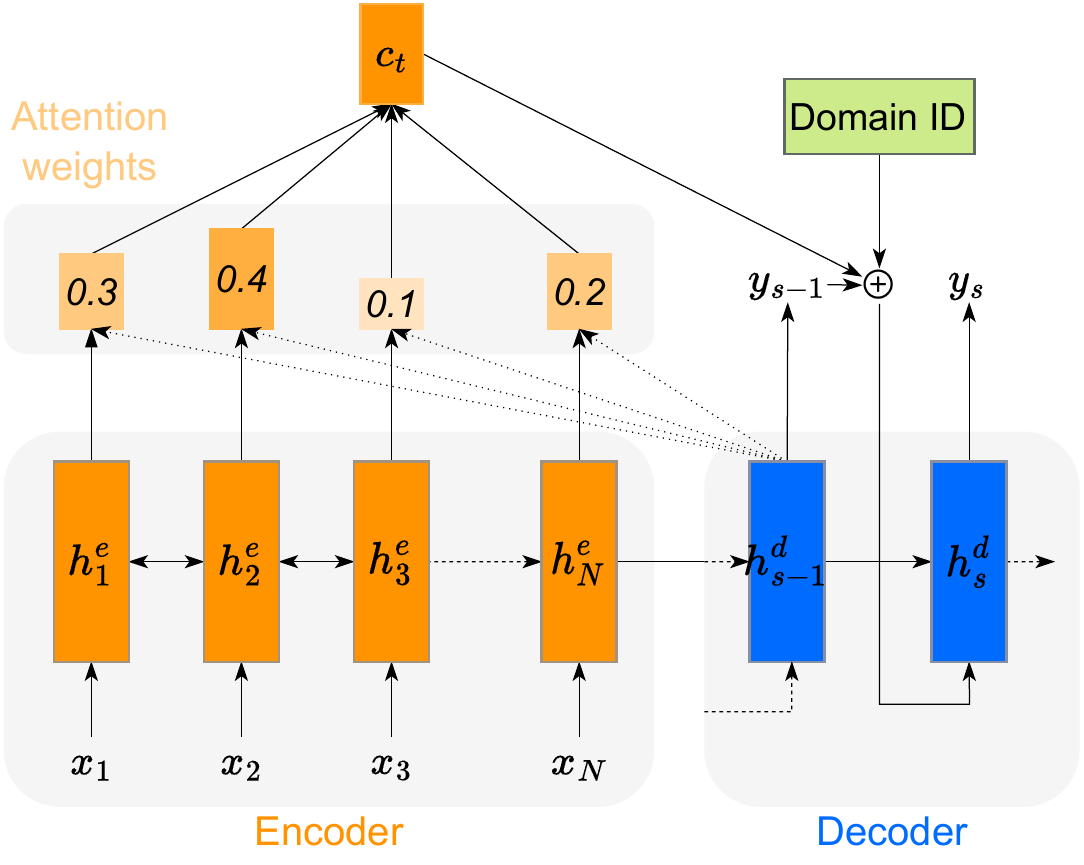}
\caption{The proposed prediction model takes pre-trained BERT word embeddings ($x_1, \dots, x_N$) and domain ID as inputs and predicts a sequence of utterance-level codes ($y_1, \dots, y_S$).}
\label{fig:arch_pred}
\end{figure}

\subsubsection{Centroid code}
\label{sec:centroid_code}

Since oracle mel-spectrograms are not available for reference during inference time, a latent vector from the learned acoustic space must be chosen differently. A common approach for the VAE latent space is to use the \emph{centroid} of a given domain by calculating the mean of all latent vectors corresponding to audio from that domain. This results in a good general representation of that domain. However, since the VQ-VAE and SVQ-VAE spaces are discrete, calculating the centroid in the same way is likely to result in a latent vector that is not represented in the codebooks and was, therefore, never observed during training. This can lead to instabilities and quality degradation. Instead, we selected a centroid code by calculating the mean of all domain vectors and performed a nearest-neighbor lookup.

\subsection{Prediction of the acoustic space from the input text}
\label{sec:pred_acoustic_space}

In order to predict the acoustic latent space, we train the model depicted in Figure~\ref{fig:arch_pred}. Rather than using a sequence of phonemes as input, the model utilizes BERT's word-piece embeddings that can capture the semantic and syntactic meaning of the text~\cite{rogers2020primer}. 

The word embeddings are consumed by the encoder’s bi-directional GRU, which further summarizes the input utterance and outputs a new representation of each word. The decoder is then conditioned on this representation to predict a code for each split $y_{s}$. $y_{1}, y_{2}, \ldots, y_{S}$ then altogether represent the sentence-level acoustic features of the utterance.
At decoding step $s$, the model uses Bahdanau-style attention~\cite{bahdanau2014neural} to obtain alignment scores for each input word, representing how important each word is for the prediction of the code $y_{s}$, which corresponds to the $s$th split. These scores are then multiplied with the corresponding hidden states of the encoder’s bi-directional GRU to produce a context vector $c_{s}$. Finally, the domain ID, the context vector $c_{s}$, the prediction from the previous step $y_{s-1}$, and the previous decoder hidden step $h_{s-1}$ are concatenated and fed into the decoder's GRU to predict the following code $y_{s}$. 

We hypothesize that autoregressive connections give the model more power to represent entangled audio features captured in codes. In addition, our experiments with non-autoregressive models often lead to a collapse, where the model predicts the same output for any input text. The Bahdanau-style attention is applied so that the model can focus on different parts of input for different splits.

\subsubsection{K-means clustering}

\begin{figure}
\centering
\includegraphics[width=0.75\linewidth]{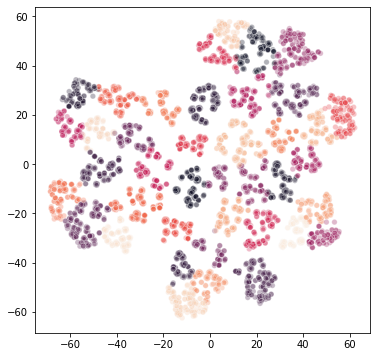}
\caption{t-SNE plot of the SVQ acoustic latent space. Different colours show the result of k-means clustering for 40 clusters.}
\label{fig:kmeans_tsne}
\end{figure}

In order to train an SVQ-VAE reference encoder that is good at representing the audio, a reasonably large number of codes and splits has to be used. Indeed, training a model with fewer splits or codes causes a deterioration in the audio quality. However, this requirement may not be necessary for the prediction task. By reducing the number of prediction targets, we can significantly simplify the training objective while still improving the naturalness and expressivity of the generated speech. As shown in Figure \ref{fig:kmeans_tsne}, visualizing the latent space with t-SNE, the codebook space forms distinct clusters. This means that we can simplify the task to the prediction of cluster centroids, significantly reducing the number of prediction targets.

We applied the K-means clustering algorithm to each codebook split and set the number of centroids to 40 using the Elbow method \cite{sugar2003finding}. Thus, we have 40 classification targets per split in our prediction task, and each cluster is represented by the closest codeword to the cluster mean.

While the k-means clustering step is not necessary for predicting the acoustic space, our internal evaluations show an improvement over the baseline model. Therefore, this approach is used in the final evaluations.


\section{Experiments}

\begin{figure*}[hbt!]
\centering
\begin{subfigure}{.4\textwidth}
  \centering
  \includegraphics[width=1.\linewidth]{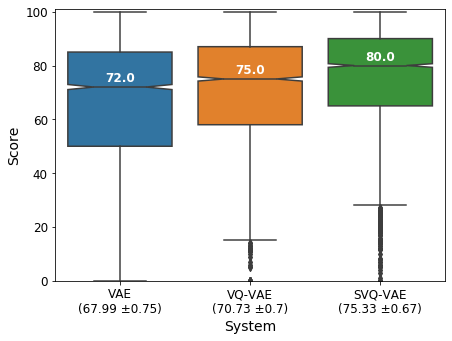}
    \caption{Centroid Study}
    \label{fig:mushra_centroid}
\end{subfigure}
\quad
\begin{subfigure}{.4\textwidth}
  \centering
  \includegraphics[width=1.\linewidth]{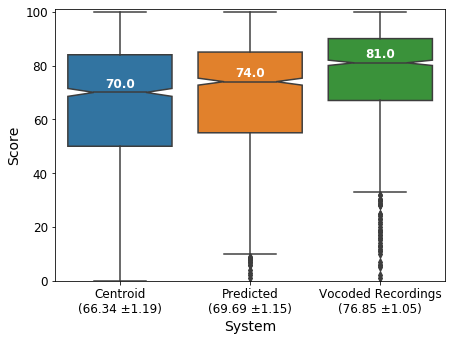}
    \caption{Prediction Study}
    \label{fig:mushra_predict}
\end{subfigure}
\caption{Results of the MUSHRA evaluations. Median scores are shown in the boxes, mean scores with 95\% confidence interval range are shown underneath.}
\label{fig:mushra_results}
\end{figure*}

In this section, we present two experiments. The first shows that the SVQ-VAE model outperforms VQ-VAE and VAE when using domain centroids for constant latent vector synthesis. The second experiment shows that the discretized space can be predicted from text, improving audio quality over synthesis with constant latent vectors. Both experiments are evaluated in the task-oriented dialogue domain.

\subsection {Voice data}

To train the speech synthesis model, we recorded scripted dialogues in English from various task-oriented domains (ticket booking, recommender systems, food orderings, etc.). A professional speaker read the turns of the agent, and a second person read the user's turns. By providing additional context to the professional speaker (i.e., including dialogue partner results), we obtained a more natural rendition of the recording script. We did not annotate the recording scripts with labels and instructed the speaker to vary their manner of speech in accordance with the context of the dialogue while maintaining a natural, conversational style. During the recordings, the speaker was also encouraged to modify the script to sound more conversational \cite{campbell2007towards}, resulting in additional contractions and the addition of filler words and pauses (e.g., \emph{OK, so, well}, \emph{Hmm}).

The number of recorded dialogues is $610$, consisting of $4,796$ agent turns, totaling $\textapprox5.5$ hours of audio. We combined this data with additional recordings from the same professional speaker: $\textapprox24$ hours of audio spoken in a neutral style in the general domain and $\textapprox50$ hours of expressive recordings in a variety of domains (news reading, emotions, etc.).

\subsection {Constant vector synthesis}
The results of the first experiment show that the SVQ-VAE model outperforms the baseline models when all utterances in the test set are synthesized using the same constant latent vector. For the constant vector, we chose the domain centroid of the recorded dialogues as a general representation of the speaking style. We used a test set of utterances consisting of $31$ dialogues with a total of $125$ turns. The set was synthesized with the VAE, VQ-VAE, and SVQ-VAE models using their respective versions of the domain centroid (see section 2.4.1). In a MUSHRA evaluation, we asked US native expert listeners to ‘Please rate the systems in terms of their naturalness,’ targeting 30 ratings per utterance. We received a total of $3,375$ ratings. The results in Figure \ref{fig:mushra_centroid} show that the SVQ-VAE model has a rating mean of $75.33$, which is higher than both the VQ-VAE and VAE models, rated $70.73$ and $67.99$, respectively. Two-sided pairwise t-tests between all model pairs show statistical significance with p-values\,$<0.01$.

\subsection {Predicted vector synthesis}

The results of the second experiment show that appropriate sampling from the latent space can benefit synthesis quality. However, deciding how to sample from the latent space for a given utterance is a challenging problem. In this work, we predict the latent space only from the text of the current utterance, leaving more complicated methods (e.g., prediction from dialogue context) to be explored in future work.

Our test set consists of $13$ task-oriented dialogues with a total of $75$ turns. It includes scenarios with successful and unsuccessful interactions in which the voice is expected to show appropriate variations (e.g., positive when a recommendation is successful).

We synthesized the set with the SVQ-VAE model using the domain centroid code described in section \ref{sec:centroid_code} and with codes predicted from text using the model described in section \ref{sec:pred_acoustic_space}. Finally, we evaluated the models in a MUSHRA test containing 3 systems: 1) SVQ-VAE using centroid; 2) SVQ-VAE with predicted codes; and 3) vocoded recordings as an upper anchor.

We asked US native expert listeners to ‘Please rate the systems in terms of their naturalness and expressivity,’ targeting $20$ ratings per utterance. In this experiment, we are interested in the overall quality of speech which includes adjusting the expressivity to match the utterance text. Therefore, we asked the listeners to additionally consider expressivity to capture this effect in the evaluation. We received a total of $1,317$ ratings. 

The results in Figure \ref{fig:mushra_predict} show that the model using predicted codes is preferred over the one using the centroid. It has a rating mean of $69.69$ compared to $66.34$, reducing the gap to the vocoded recordings by $32\%$. Two-sided pairwise t-tests between all model pairs show statistical significance with p-values\,$<0.01$.


\section{Conclusion}

We introduced SVQ-VAE, a novel NTTS architecture that allows modeling a discrete space of global prosodic representations. We applied it to the challenging domain of task-oriented dialogues. We demonstrated that our architecture outperforms standard VAE and VQ-VAE baselines when inferring with a centroid. Additionally, we proposed an autoregressive predictor model that utilises contextualised representation from BERT embeddings to predict SVQ-VAE representations to further improve our system. Inferring with predicted representations outperformed centroid-based synthesis and reduced the gap to vocoded recordings by $32\%$ in terms of average MUSHRA score. For future work, we plan to improve the prediction of the latent space by utilizing additional contextual features (e.g., dialogue embeddings), addressing the appropriateness of multi-turn dialogues rather than individual utterances.

\vspace*{\fill}
\pagebreak

\bibliographystyle{ieeetr}
\bibliography{ms}

\end{document}